\documentclass[preprint,showpacs,preprintnumbers,amsmath,amssymb]{revtex4}

\usepackage{graphicx}
\usepackage{dcolumn}
\usepackage{bm}

\begin{document}


\title{Phenomenological understanding of\\
 aggregation and dispersion of chemotactic cells}

\author{Masatomo Iwasa}
 \email{miwasa@r.phys.nagoya-u.ac.jp}
\author{Ryosuke Ishiwata}%
 \altaffiliation[Also at ]{Department of Bioinformatics, Graduate School
 of Medicine and Dentistry, Tokyo Medical and Dental University, Tokyo
 113-8501, Japan}
\affiliation{%
Department of Complex Systems Science,
 Graduate School of Information Science, Nagoya University, Nagoya
 464-8601, Japan}%

\date{\today}

\begin{abstract}
 We present a simple model that describes the motion of a
 single chemotactic cell exposed to a traveling wave of the
 chemoattractant.
The model incorporates two types of responses to stimulation by the
 chemoattractant, i.e., change in polarity and change in motility of the cell.
The periodic change in motility is assumed to be induced by the periodic
 stimulation by the chemoattractant on the basis of previous
 observations. 
Consequently, net migration of the cell occurs in a particular direction
 with respect to wave propagation, which explains the migration of {\it
 Dictyostelium} cells in aggregation processes.
The difference between two time delays from the stimulation to the two
 responses and the wave frequency determined by the frequency of the
 secretion of the chemoattractant are important parameters that
 determine the direction of migration and the effective interaction
 between cells in a population. 
This result explains the dispersed state of a population of vegetative
 cells and cells in preaggregation without the assumption of a
 chemorepellent, and also explains the commencement of the aggregation.  
The result is extended to a general fact as follows: 
 the temporal oscillation of the magnitude of the random motion for
 gradient-sensing particles induces spontaneous movement, even when the
 particles are exposed to a periodic wave of the chemoattractant, which
 results in the aggregation or dispersion of the particles communicating
 via their attractant. 
\end{abstract}

\pacs{Valid PACS appear here}
\maketitle
\section{Introduction}
Several types of cells can sense the presence of extracellular signals
and migrate in the direction of the concentration gradient of these
signals.  
This phenomenon is known as chemotaxis, which plays an important role
in a variety of biological systems including mammalian neutrophilis
\cite{Stephens2008}, fibroblasts \cite{Schneider2006}, microglia
\cite{Rogers2001}, and cancerous cells \cite{Kedrin2007}. 
{\it Dictyostelium discoideum} serves as an ideal model for the study of
chemotaxis. 
Each {\it Dictyostelium} cell migrates toward the gradient of the
chemoattractant, {\it cyclic adenosine monophosphate} (cAMP) 
\cite{Song2006}.
It is well known that a population of {\it Dictyostelium} cell exhibits
collective behavior. 
When bacterial food is available, {\it Dictyostelium} cells live as
unicellular amoebae.  
In the absence of food, the developmental phase of the life cycle is
induced, that is, the cells aggregate into a multicellular slug, and
form a fruiting body whose spores germinate into amoebae. 
In the aggregation process, the dispersed starved cells periodically
emit extracellular pulses of the chemoattractant 
\cite{Tomchik1981, Devreotes1983, Gregor2010}, 
a target or rotating wave of the chemoattractant concentration is
formed, 
and the cells migrate toward the center of the wave via chemotaxis
\cite{Song2006, Lee1996}. 

In this article, we present a highly simplified phenomenological model
 that describes the motion of a single chemotactic cell.
The analytical results allow us to explain the migration of cells
 toward the wave source in the aggregation process and to give an
 explanation for the aggregation of starved cells and dispersed state of
 vegetative cells and preaggregating cells in a population of {\it
 Dictyostelium} cells.   

Simplified and generalized descriptions of chemotactic migration
are desired not only for the understanding of natural sciences but also
for various industrial applications. 
Recently, numerous phenomena that emerge through natural
self-organization processes have inspired industrial applications such 
as nanofabrication \cite{Whitesides2002} and robotics 
\cite{Ishiguro2006, Pfeifer2007}. 
{\it Dictyostelium} is a typical example of gradient-sensing systems
that facilitate self-organization of the population; therefore, 
the understanding of the mechanism as simply as possible would
contribute to the development of artificial systems and related
technologies.  
The mechanism of migration presented in this study is sufficiently
 simple to allow application of the mechanism to other gradient-sensing
 systems. 
One such system is presented in the last section.

\section{Model}
\begin{figure}[t!]\vspace*{3pt}
\centering{\includegraphics[scale=0.9]{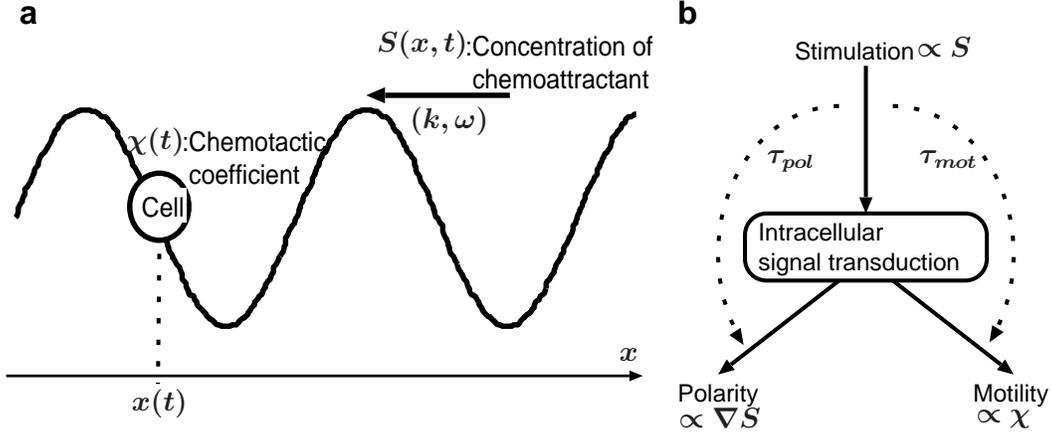}}\vspace*{-6pt}
\caption{
(a) The situation considered in this study. 
 A single chemotactic cell is exposed to a chemoattractant wave
 propagating from right to left. The cell senses the gradient of the 
 chemoattractant and the temporal change in the concentration.
(b) Illustration which schematically shows the construction of the 
 model. 
 The stimulation by the traveling wave of the chemoattractant induces
 two responses, change in polarity and motility, with time delays,
 $\tau_{pol}$ and $\tau_{mot}$.  
\label{fig:1}}\vspace*{-8pt} 
\end{figure}

Consider a single cell in one dimension, exposed to a traveling wave of
 the chemoattractant concentration (see Fig. \ref{fig:1}(a)) represented
 by a sinusoidal function:
\begin{eqnarray}
 S(x,t)=S_{ave}+S_{osc}\sin\left(kx+\omega t\right),
\label{eq:2-1}
\end{eqnarray}
at position $x$ and time $t$.
Here, $k$ and $\omega$ denote the wave number and the angular frequency,
respectively.  
$S_{ave}$ and $S_{osc}$ are constant parameters which satisfy
 $S_{osc}<S_{ave}$.
Although it has been observed that the typical waveform is not a simple
sinusoidal form but a sharply peaked one for a population of 
{\it Dictyostelium} cells\cite{Tomchik1981}, as we discuss later, the
use of the sinusoid is not essential to the result.  
The result qualitatively holds as long as the wave is monophasic and
periodic. 

The model presented in this study is such that the position of a single
 cell, $x(t)$, is governed by the following equation of motion,
\begin{eqnarray}
\frac{{\rm d}}{{\rm d}t}x(t)
 &=&\chi(t) \nabla S(x(t-\tau_{pol}),t-\tau_{pol}), 
\label{eq:2-2}\\
    \chi(t)\hspace{-1mm}
    &:=&\chi_{ave}\hspace{-1mm}+\hspace{-1mm}\chi_{osc}
          \sin[kx(t-\tau_{mot})\hspace{-1mm}
                              +\hspace{-1mm}\omega(t-\tau_{mot})]. 
\label{eq:2-3}
\end{eqnarray}
Here $\chi_{ave}$, $\chi_{osc}$, $\tau_{mot}$ and $\tau_{pol}$ are
constant parameters which satisfy $\chi_{osc}<\chi_{ave}$.

This model equation is constructed as follows (Fig. \ref{fig:1}(b)).
We use a widely-accepted assumption that the velocity of the cell is
 proportional to the spatial gradient of the chemoattractant 
 concentration \cite{Keller1970,Othmer1998,Goldstein1995}. 
The proportional coefficient, $\chi$, is referred to as chemotactic
coefficient in what follows. 
Recent studies on intracellular signal transduction pathways of
chemotactic cells have shown that the stimulation by extracellular
chemoattractant triggers three responses which should be separately
recognized, namely, directional sensing, polarization, and change in
motility 
\cite{FrancaKoh2006, Iglesias2008, Swaney2010}.   
In terms of those studies, from the viewpoint of the modeling of the
motion of a cell, we assume that the responses induced by the
stimulation by a traveling wave consequently can be divided into two
responses.
One of them is the set of the directional sensing and the polarization,
that is, the spatial localization of several signaling proteins along
the plasma membrane because of the presence of the spatial gradient of
the chemoattractant concentration in the environment, which results in
pseudopod extension to a particular direction. 
We assume that the factor proportional to $\nabla S$ describes the
magnitude of this response.  
The constant parameter, $\tau_{pol}$, is introduced as the time required
from the stimulation to this response. 
Another response is the change in motility, namely, the magnitude and 
frequency of the extension of pseudopodia.
In the observation of cell locomotion, the motility would be recognized
as the quantity which indicates how fast a cell translocates on the
substrate.   
Thus, we assume that the chemotactic coefficient, $\chi$, is the factor
which quantify the level of this response; therefore, it is assumed that
the chemotactic coefficient oscillates in response to the oscillation of
$S$.  
The constant parameter, $\tau_{mot}$, is introduced as the time required
from the stimulation to this response. 

The assumption of the oscillatory behavior for the chemotactic
coefficient is also based on remarkable previous studies on cell
locomotion. 
Soll et al. determined the time series of the instantaneous velocity of
a single {\it Dictyostelium} cell in detail when the cell was exposed to
temporally periodic change in the chemoattractant concentration in the
absence of the spatial gradient \cite{Soll2002, Zhang2002, Zhang2003}.  
It was observed that the magnitude of the instantaneous velocity
oscillated at the same frequency as the extracellular chemoattractant
concentration.  
This result implies that the temporal change of the chemoattractant
concentration induces the change in the magnitude of the motility,
which means here how fast a cell translocates.
When a cell is exposed to a periodic traveling wave of the
chemoattractant concentration, 
there is also the oscillation of extracellular chemoattractant
concentration around a cell. 
Therefore, the model assumes that the chemotactic coefficient
oscillates.  
In addition, because a significant constant phase difference was
observed between the oscillation of the extracellular chemoattractant
concentration and that of the instantaneous velocity in the experiments,
the time delay, $\tau_{mot}$, corresponding to the phase difference is
introduced in the model.   

The assumption in this model provided above can be described in another 
way from the theoretical point of view.  
The equation of motion of a single cell is generally given by
\begin{eqnarray}
 \gamma\frac{d}{dt}x&=&F+\xi,
\label{eq:2-4}
\end{eqnarray}
as long as we neglect the inertial effect of a cell.
Here, $F$ denotes external forces, $\xi$ denotes forces which
induces random locomotion of a cell \cite{Li2008,Takagi2008}, and the
coefficient $\gamma$ is referred to as the friction coefficient.  
The equation reads
\begin{eqnarray}
       \frac{d}{dt}x&=&\gamma^{-1}F+\gamma^{-1}\xi.
\label{eq:2-5}
\end{eqnarray}
In the absence of the spatial gradient of the chemoattractant and other
external forces, the governing equation becomes
\begin{eqnarray}
       \frac{d}{dt}x&=&\gamma^{-1}\xi.
\label{eq:2-6}
\end{eqnarray}
According to the experiments carried out by Soll et al.
\cite{Soll2002, Zhang2002, Zhang2003}, the temporally periodic
and spatially uniform stimulation by the chemoattractant induces the
periodic change in the instantaneous velocity, $\frac{d}{dt}x$, at the
same frequency with a constant time delay.  
Here, we attribute the oscillation of the velocity to the oscillation of
$\gamma^{-1}$. 
On the other hand, when a cell is exposed to a traveling wave of the
chemoattractant concentration $S$, $F$ is proportional to $\nabla S$,
that is, 
\begin{eqnarray}
 \gamma\frac{d}{dt}x=\nabla S+\xi.
\label{eq:2-7}
\end{eqnarray}
Omitting the random component $\xi$ since we are interested in the
time-averaged quantities in this study, 
Eq. (\ref{eq:2-4}) is reduced to
\begin{eqnarray}
 \frac{d}{dt}x=\chi\nabla S,
\label{eq:2-8}
\end{eqnarray}
where $\chi$ is proportional to $\gamma^{-1}$.
Because the magnitude of the extracellular chemoattractant concentration
periodically changes when the cell is exposed to a periodic traveling
wave of the chemoattractant, we use the results obtained by Soll et
al., that is, 
we assume that the chemotactic coefficient $\chi$ oscillates at the
same frequency as the wave.
Incorporating the time delays to the responses, we obtain the model
equation (\ref{eq:2-2}) in consequence.
Although the periodic function that precisely describes the chemotactic
coefficient has never been determined in the presence of the spatial
gradient of the chemoattractant concentration, we choose a sinusoidal
function for simplicity.  
As discussed later, this choice of the sinusoidal function is not
essential to the result; the result holds as long as the chemotactic
coefficient is periodic function.  

In summary, we postulate in the modeling that a response, change in how 
large or frequently pseudopodia are extended, can be distinguished  
from a response, determination of the direction of locomotion; 
the level of the pseudopod extension is affected by the temporal change
in the magnitude of the extracellular chemoattractant concentration 
whereas the direction decision is affected by the spatial gradient of the 
extracellular chemoattractant concentration; and the composition of
these factors determines the instantaneous velocity of a cell exposed to
a traveling wave.
Explicit incorporation of the two time delays, $\tau_{pol}$ and
$\tau_{mot}$, is an essential feature of this model.
It should be noted that this model is not another expression of models
presented thus far in which the periodic change in motility is
considered  but an essentially different model which provides another
result on cell migration.   
The significant difference between them is discussed later.
\begin{table*}[bhtp!]
\begin{tabular}{@{\vrule width 1pt}l|l|r|r@{\ \vrule width 1pt}}
 \hline
 Abbreviation & Meaning & Used value & Ref. \\ \hline
 $2\pi/k$ & Wave length of the chemoattractant wave 
   & 2000 [$\mu m$] & \cite{Muller1998} \\ \hline 
 $2\pi/\omega$ & Periodic time of the chemoattractant wave 
   & 7 [$min$] & \cite{Muller1998} \\ \hline 
 $k\chi_{ave}S_{osc}$ & Averaged velocity of the single cell
   & 10[$\mu m/min$] & \cite{Steinbock1991}\\ \hline 
 $k\chi_{osc}S_{osc}$ & Amplitude of the oscillation of the velocity of
 the single cell 
   & 10[$\mu m/min$] & \cite{Steinbock1991}\\ \hline 
 $\tau_{mot}$ & time delay from the stimulation to the change in
 motility
   & 6[$min$] & \cite{Soll2002} \\ \hline
 $\tau_{pol}$ & time delay from the stimulation to the change in
 polarity
   & 1[$min$] & * \\ \hline
\end{tabular}
\caption{
Model parameters used in the numerical calculations whose results are
 shown in Fig. 2.
All of them are typical values for {\it Dictyostelium} cells in
 aggregation process except for $\tau_{mot}$. 
Although $\tau_{pol}$ has not been measured thus far, 
it is estimated with the model in this study (see the Discussion in the
 main text).
\label{table:1}}
\end{table*}

\section{Result}
We immediately see that the propagation of the traveling wave induces
net migration of a cell to a particular direction in average.  
We can safely approximate $x(t-\tau_{pol})=x(t-\tau_{pol})=x(t)$
because the velocity of the traveling wave, $\omega/k\sim300[\mu m/min]$
\cite{Tomchik1981}, 
is much greater than that of a cell, which varies between $0$
and $20[\mu m/min]$ \cite{Soll2002}. 
Thus, the velocity of the single cell is governed by
\begin{eqnarray}
 \frac{{\rm d}}{{\rm d}t}x(t)
       &=&k\chi_{ave}S_{osc}
           \cos\left[kx(t)+\omega(t-\tau_{pol})\right]
\nonumber\\
        &&+(k\chi_{osc}S_{osc}/2)
       \sin\left[2kx(t)+\omega(2t-\tau_{mot}-\tau_{pol})\right]
\nonumber\\
        &&-(k\chi_{osc} S_{osc}/2)
             \sin\left[\omega(\tau_{mot}-\tau_{pol})\right].
\label{eq:3-1}
\end{eqnarray}
On the right hand side of this equation, the third term is a constant
 whereas the first and second terms are periodic with respect to time as
 long as $x(t)$ is treated as a constant during one wave period.
Therefore, the third term plays a greater role in the translocation of a
 cell than the first and second terms.
The elimination of these periodic terms yields the averaged velocity
 of the cell, $\bar{v}$, as 
\begin{eqnarray}
 \bar{v}=-(k\chi_{osc} S_{osc}/2)
             \sin\left[\omega(\tau_{mot}-\tau_{pol})\right].
\label{eq:3-2}
\end{eqnarray}

\begin{figure}[b!]\vspace*{3pt}
\centering{\includegraphics[scale=1.0]{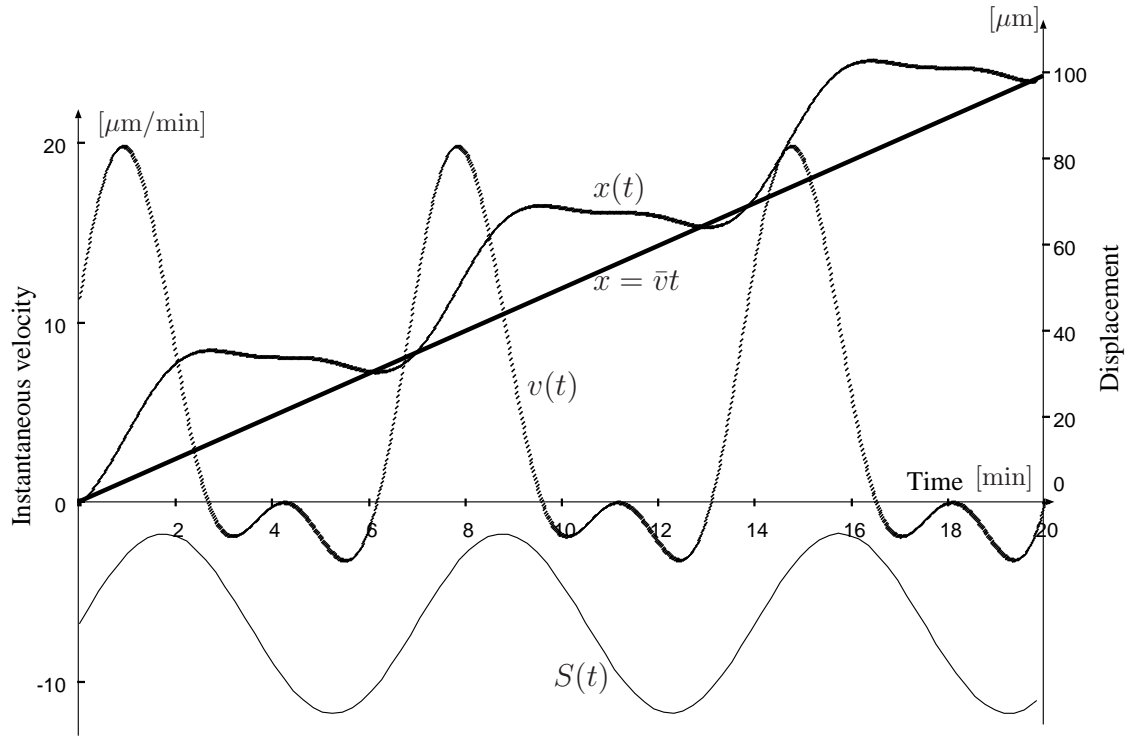}}\vspace*{-0.0pt}
\caption{The time course of the velocity $v(t)$ and the displacement of a
 cell $x(t)$ obtained by numerically solving the differential equation 
 (\ref{eq:2-2}) without any approximation.
The thin curved line at the bottom $S(t)$ represents the chemoattractant 
 concentration, which is shown as a reference in order to be compared
 with the phase in the oscillation of the velocity and the displacement. 
The straight line represents the approximate displacement $\bar{v}t$
 calculated with the averaged velocity $\bar{v}$ in Eq. (\ref{eq:3-2}).
\label{fig:2}}\vspace*{-8pt} 
\end{figure}

We see that the cell migrates toward or away from the wave source on
average when  
$\sin \left[ \omega (\tau_{mot}-\tau_{pol}) \right] < 0$ or 
$\sin \left[ \omega (\tau_{mot}-\tau_{pol}) \right] > 0$, respectively. 
Thus, the frequency of the wave and the difference between the two time 
delays, $\tau_{mot}-\tau_{pol}$, are important parameters that
determine the direction and the magnitude of the net migration. 

The differential equation (\ref{eq:2-2}) is numerically solved without
any approximation. 
The result is illustrated in Fig. \ref{fig:2}.
Parameter values used here are summarized in Table \ref{table:1}. 
We see that the averaged velocity well describes the net migration of a
cell. 

Even if the waveform is not represented by a simple sinusoidal
function, $\omega$ and $\tau_{mot}-\tau_{pol}$ play important roles in
determining the direction of the net migration.  
Consider a more realistic case in which the traveling wave of the
chemoattractant concentration is a sharply peaked one represented by
\begin{eqnarray}
S(x,t)=S_0\exp[S_1\sin(kx+\omega t)], 
\label{eq:3-3}
\end{eqnarray}
where $S_0$ and $S_1$ are constant parameters, and the chemotactic
coefficient is given by Eq. (\ref{eq:2-3}).
These functional forms of $S(x,t)$ and $\chi(t)$ describe the result of 
the experiment conducted by Soll et al. more precisely 
\cite{Soll2002, Zhang2003, Zhang2002}. 
In the same manner as the case of the sinusoidal traveling wave, it is 
proved by means of the Fourier expansion that   
\begin{eqnarray}
 \bar{v}=-(\chi_{osc}S_0S_1kI/\pi)\sin[\omega(\tau_{mot}-\tau_{pol})],
\label{eq:3-4}
\end{eqnarray}
where $I:=\int_{-1}^{1}dt\sqrt{1-t^2}\exp(S_1t)>0$.
We see that the direction of net the migration of a cell is determined
by the sign of $\sin[\omega(\tau_{mot}-\tau_{pol})]$; 
therefore, the the frequency of the traveling wave and the difference of
the two time delays are important parameters again.

\begin{figure}
 \begin{center}
  \includegraphics[scale=1.2]{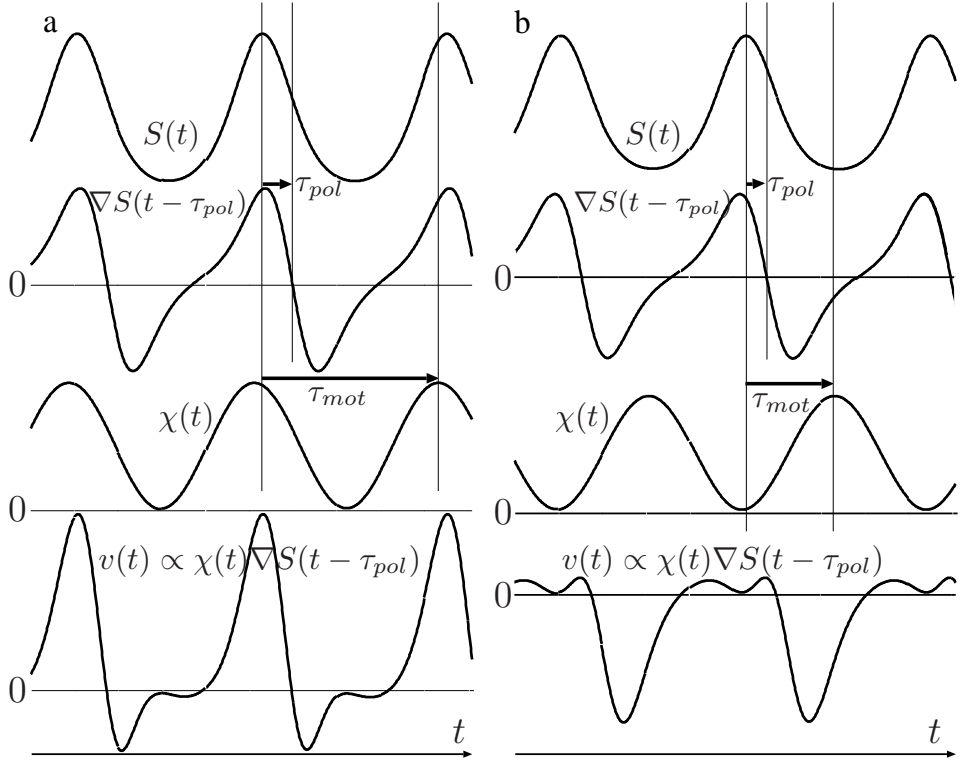}
 \end{center}
\caption{
(a) The relationship among the extracellular chemoattractant
 concentration $S$, its gradient with a time delay $\nabla
 S(t-\tau_{pol})$, the chemotactic coefficient with a time delay
 $\chi(t)>0$, and the velocity $v(t)\propto\chi(t)\nabla
 S(t-\tau_{pol})$, in the case which results in net migration toward the
 source of the traveling wave.
(b) The case which results in net migration away from the source of the
 traveling wave. 
The Figure (a) and the Figure (b) also illustrate the situations in
 which the frequency of the traveling wave is small and large,
 respectively.    
Distinction of these situations is important not only to see the
 direction of the net migration of cells but also to infer the effective 
 interaction between cells (see the Discussion for detail). 
\label{fig:3}
}
\end{figure}

More generally, let us discuss the reason for the emergence of the net
migration with referring to Fig. \ref{fig:3}.
If the shape of the wave of the chemoattractant concentration $S$
is monophasic and periodic, its spatial gradient $\nabla S$ becomes a
biphasic wave whose time average is 0.
If the time delays are such that the peak response of the oscillating
motility $\chi(t)$ appears when $\nabla S(t-\tau_{pol})$
is positive, the time average of the product of $\chi(t)$ and $\nabla
S(t-\tau_{pol})$, which is proportional to $\bar{v}$, becomes positive.
This implies that the cells migrate toward the source of the traveling
wave (Fig. \ref{fig:3}a).  
Conversely, if the time delays are such that the peak response of the
motility $\chi(t)$ appears when $\nabla S(t-\tau_{pol})$ is negative,
the time average of the product of $\chi(t)$ and $\nabla
S(t-\tau_{pol})$ becomes negative.
This implies that cells migrate in the direction of the wave propagation   
(Fig. \ref{fig:3}b). 
Thus, even if the wave is not represented by a simple sinusoidal
function, 
the difference between the two time delays from the stimulation to the
two responses determines the direction of the cell migration of the
cells.  
Moreover, as long as we fix the two time delays, $\tau_{mot}$ and 
$\tau_{pol}$, Fig. \ref{fig:4}a and Fig. \ref{fig:4}b illustrate
situations in which the frequency of the traveling wave, $\omega$, is
small and large, respectively.  
This indicates that the direction to which a cell migrates also depends
on the wave frequency $\omega$. 
Thus, the mechanism of directional migration of a cell is fundamentally
the same as long as a wave is monophasic and periodic. 
The frequency of the traveling wave and the difference of the two time
delays are important parameters which determine the direction of the
net migration. 

Note that, if there are no such time delays, that is, if
$\tau_{mot}=\tau_{pol}=0$, the time-average of $\chi\nabla S$ becomes 
0 independently of the functional form of $S$. 
The directional net migration of cells has been successfully derived
 using several models in which time dependence of chemotactic
 coefficient is taken into account 
\cite{Goldstein1995, Hofer1994, Vasiev1994, Oss1996, Dolak2005}.  
On the basis of the discussion provided above, we can conclude that all
these models have been constructed such that the time delays are
implicitly incorporated in order to introduce an appropriate phase
difference between the oscillation of the stimulation and the responses.

\section{Discussion}

This result (\ref{eq:3-2}) and (\ref{eq:3-4}) are consistent with the
aggregation of {\it Dictyostelium} cells in that the each averaged
velocity $\bar{v}$ is positive for typical parameter values in the
aggregation process shown in Table \ref{table:1}, which indicates that
cells migrate toward the center of a target or spiral wave of cAMP
concentration. 

The numerical result shown in Fig. \ref{fig:2} well duplicates previous
observations of the instantaneous velocity of a single 
{\it Dictyostelium} cell during the aggregation processes 
\cite{Soll2002, Zhang2002, Zhang2003} in that a period in which a 
cell migrates rapidly alternates with a period in which a cell is
virtually stationary.   
In particular, the result successfully reproduces the observation more
precisely than another model \cite{Goldstein1995} in that the time
courses obtained in the experiments displays complex behavior which
cannot be constructed by a simple sinusoid such as the characteristic
small peaks in the stationary period.

The result provided above implies that there exists an effective
interaction between cells, because, in the case of {\it Dictyostelium},
each cell secretes the chemoattractant, the secreted chemoattractant
propagates in space, and it serves as a stimulation to nearby cells
\cite{Gregor2010}. 
More precisely, if the frequency of the secretion of the
chemoattractant, which determines the wave frequency, and the time
delays satisfy the condition that a cell migrates toward or away from
the wave source, there is effective attraction or repulsion between
cells, respectively.   
This fact indicates that, for systems in which cells communicate with
each other using the chemoattractant as an autoinducer, the frequency of
the secretion and the time delays determine the effective interaction
between the cells. 
In aggregation process, because the parameter values satisfy
$\bar{v}>0$, the effective interaction between cells is attraction.
This attraction is interpreted as the microscopic reason for the
aggregation of cells. 

Increase of the frequency of the secretion of the chemoattractant
during aggregation has been observed \cite{Gregor2010, Alcantara1974}.  
The averaged velocity (\ref{eq:3-2}) and (\ref{eq:3-4}) indicate that,
as the frequency of the secretion of the chemoattractant, $\omega$,
increases, the sign of $\bar{v}$ changes from negative to positive at a
critical frequency $\omega_c$, which implies that the effective
interaction between cells changes from repulsion to attraction.  
Even if the wave form is not so simple as Eq. (\ref{eq:2-1}) or
Eq. (\ref{eq:3-3}), we can infer that the transition occurs according to
the discussion provided above and Fig. \ref{fig:3}.  
Therefore, it is possible that the dispersed state of cells before the
commencement of aggregation is due to the effective repulsion between
cells induced by the lower frequency of the secretion of the
chemoattractant. 

In terms of the transition between the dispersion and aggregation in a
population of cells, we can estimate $\tau_{pol}$ for 
{\it Dictyostelium} cells. 
From Eq. (\ref{eq:3-2}) and Eq. (\ref{eq:3-4}), the frequency of the
secretion of the chemoattractant at which the interaction between cells
changes from repulsion to attraction, $\omega_c$, satisfies 
$\tau_{mot}-\tau_{pol} = \pi/\omega_c$.  
According to a previous experiment, the time period of the periodic
secretion at the commencement of the aggregation was 
$2\pi/\omega_c\sim 10$ [min] \cite{Alcantara1974}.  
Then, since $\tau_{mot}\sim 6$ [min] \cite{Soll2002}, we obtain
$\tau_{pol}\sim 1$ [min]. 
Although this parameter $\tau_{pol}$ has not been directly measured in
the presence of temporally periodic spatial gradient of the
chemoattractant concentration thus far, the result is consistent with
an estimation based on the time which requires for the localization of
proteins on the plasma membrane \cite{Rappel2009}. 
Inversely, this consistency suggests the existence of the transition at
around $2\pi/\omega_c\sim 10$ [min].

Repulsion between {\it Dictyostelium} cells in a nutrient environment
was reported several decades ago \cite{Samuel1961, Keating1977}.
Although experimental studies on vegetative cells are less than starved
cells in the developmental phase, it is certain that vegetative cells
are single-celled amoeba and never commence the aggregation.
The present model provides us an explanation for this state. 
It has been observed that right after the nutrient deprivation (within 5
hours precisely), cells pulse chemoattractant randomly every 15 to 30
min \cite{Gregor2010}. 
It can be inferred that vegetative cells secrete the cAMP at the same or
lower frequency.  
According to the discussion above and in terms of Fig. \ref{fig:3},
the interaction between cells is repulsion at those frequencies. 
Note that, even if the wave consists of a single pulse, there arise also
repulsion. 
Thus, it is possible that the effective repulsion mediated by the
chemoattractant allows vegetative cells to live as unicellular amoebae
and never start to aggregate.   

The fact that the interaction between cells may become repulsion
 indicates the possibility of the non-existence of the chemorepellent.
Repulsion between vegetative {\it Dictyostelium} cells was originally
 named negative chemotaxis \cite{Keating1977}.
Negative chemotaxis was later defined as a variant of chemotaxis in
which cells migrate away from higher concentrations of a chemical
source. 
Such a chemical substance is called chemorepellent.
Due to this historical background, it was believed that vegetative cells
 secrete a chemorepellent and repel each other owing to the
 chemorepellent. 
A potential candidate for the chemorepellent was reported recently
\cite{Keizer-Gunnink2007}.
However, on the basis of the present model, such a chemorepellent is not
 required to explain the repulsion between cells. 
Even if there is no chemorepellent, and cells exhibit only positive
chemotaxis, that is, $\chi$ is always positive, attraction and repulsion
can be switched depending on the frequency of the secretion of the
 chemoattractant.
We should note that this discussion just gives an possible scenario
 which explains for behavior of a population of vegetative cells because
 it has been commonly recognized and accepted that the intracellular
 molecular dynamics such as signal transduction and gene expression is
 largely different between vegetative cells and starved cells. 
The response of vegetative cells to cAMP should be investigated more in
 the future in order to confirm the discussion here.

It seems that the model presented here is just an another expression of
what has been presented thus far in terms of time delay or phase
difference of oscillations.
However, there is an essential difference between them. 
Although it has been also assumed that the motility of cells under a
traveling wave of the chemoattractant changes, 
the motility is believed to increase during periods in which the
concentration of the external chemoattractant increases, and decrease
during periods in which the concentration of the external
chemoattractant decreases \cite{Soll2002}. 
This has been considered as the reason why the traveling-wave chemotaxis
occurs.   
However, on the basis of this conventional idea, a cell exposed to a
traveling wave always migrate toward the wave source independently of
the frequency of the wave; therefore the effective interaction between  
cells becomes always attraction according to the discussion presented
above.
This does not explain for the transition from the dispersed state
to the aggregation of cells.  
It should be experimentally confirmed in the future that to which
direction cells migrate when cells are exposed to a traveling wave
whose frequency is much lower than the wave emerging in aggregation
processes.  
Note that in such an experiment the wave is not necessarily a periodic
wave if the durations between peaks are long sufficiently.
If those cells migrate away from the wave source, the model presented in
this study would be valid.  
Or this model will be supported if it is observed that cells right
before the commencement of aggregation significantly migrate in the
direction of the wave propagation. 

\section{Concluding Remarks}
A phenomenological model which describes the motion of a single
chemotactic cell is presented and analyzed in this study. 
The result of the analysis implies that systems in which cells
communicate with each other using the chemoattractant as an autoinducer
exhibit aggregation and dispersion of cells.
The frequency of the secretion of the chemoattractant and the time
delays from the stimulation by the chemoattractant to the responses of
polarity and motility determine the interaction between cells,
attraction or repulsion.   

It must be controversial whether or not the responses to the stimulation
by the chemoattractant can be incorporated as simply as this model
because the intracellular signal transduction consists of actually
complicated processes. 
However, the incorporation allows us 
1) to understand vegetative cells living as unicellular amoebae and not
starting aggregation in terms of repulsion between cells, 
2) to understand commencement of the aggregation of starved cells in
terms of the transition from the repulsion to attraction, 
3) to derive the cell migration toward the source of the target or
spiral wave of the chemoattractant in aggregation,
4) to describe the time course of the velocity of a cell in aggregation
more precisely that other models presented thus far, and 
5) to estimate the time which requires for the polarization.
 
The model in this study just provides at present a possible mechanism
which is used in cell migration and induces dispersion and aggregation
of cells.   
The validity of this model should be confirmed by some
experiments in the future. 
1) $\tau_{mot}$ and $\tau_{pol}$ are regarded as 
constant in this study. 
It should be confirmed whether or not values of $\tau_{mot}$ and
$\tau_{pol}$ depend on the frequency of the stimulation, and taking
the dependence into consideration also leads to the same results. 
Note that, even if those parameters depend on the frequency, the
mechanism proposed in this study works as long as there remains the
critical frequency at which the transition occurs.
2) We have used the value of $\tau_{mot}$ that was measured in the
absence of spatial gradient of cAMP in this study. 
Whether or not the change in $\tau_{mot}$ is the same in the presence of
the spatial gradient of cAMP.
3) Because it has been reported that many of responses occur within a
few minutes \cite{FrancaKoh2006}, why the change in motility requires
such a long time, about $6 [min]$, should be elucidated in terms of the
intracellular dynamics.   
4) Net migration of cells exposed to a traveling wave whose frequency is
small should be investigated in order to compare with models presented
thus far (see the preceding section for detail).
As long as we refer to the results obtained in this study, the
intriguing question arising next is how cells regulate the frequency of
the secretion of chemoattractant depending on the environment, that is,
how they choose sufficiently low frequency in order to live as
unicellular amoebae when they can survive and grow thanks to enough
food, and how they choose sufficiently high frequency in order to start
development when they cannot survive alone because of starvation.  
The answer to this question will elucidate the mechanism of the
adaptation of cells to their environment and the strategy taken for the
species survival.  
 
From the viewpoint of physics, it should be remarked that the random
motion of a cell plays an important role in the generation of
directional migration even though the random motion itself is
nondirectional, and does not seem to contribute to tactic behavior.    
In addition, it should be emphasized that the composition of symmetric
conditions induces asymmetric behavior, that is, the composition of the
periodic stimulation and the nondirectional fluctuation of the position
generates directional net migration.   
This spontaneous symmetry breaking leads to aggregation or dispersion in
 a population. 

\begin{figure}[t!]\vspace*{3pt}
\centering{\includegraphics[scale=1.0]{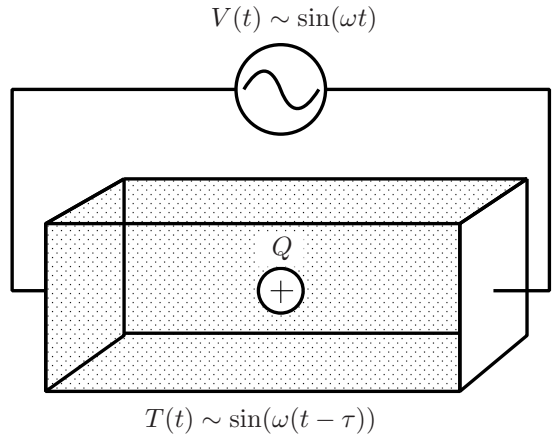}}\vspace*{-5.5pt}
\caption{
An example in which directional motion occurs with the same mechanism
 described by the proposed model.  
Consider an electrically charged particle in a solution whose
 temperature can be controlled.
 The particle is sufficiently small to move randomly in terms of the
 thermal fluctuation.
If the particle is exposed to an AC electrical field 
and if the temperature is changed with the same period as the AC field
 with a constant time delay $\tau$, the particle moves in a particular
 direction on average. 
The time delay $\tau$ determines the direction of movement.
\label{fig:4}}\vspace*{-8pt}
\end{figure}

Let us further generalize the results.
Consider an isolated element that moves in response to the spatial
gradient of a field.
When such an element is exposed to the temporally periodic change 
of spatial gradient of the field and the motility oscillates in the same 
frequency, spontaneous directional motion occurs on average.  
The direction of the motion is determined by the frequency of the change
in the gradient and the phase difference between the oscillation of the
motility and that of the response to the spatial gradient
(Fig. \ref{fig:4}). 
Moreover, a group of elements, each of which emits a field with the same
frequency (e.g., quantum dots exposed to electromagnetic waves
\cite{Iida2006}), is expected to exhibit aggregation or dispersion as
seen in {\it Dictyostelium} cells. 
In the future, such a simple mechanism can potentially facilitate the
development of self-organizing systems in which elements communicate or
interact with each other via a field.
\vspace*{-3mm}

\begin{acknowledgments}
The authors would like to thank Dr. Dan Tanaka for fruitful discussions. 
\end{acknowledgments}


\end{document}